\documentstyle[12pt]{article}
\newcommand{\be}{\begin{equation}}
\newcommand{\ee}{\end{equation}}
\newcommand{\ba}{\begin{eqnarray}}
\newcommand{\ea}{\end{eqnarray}}
\newcommand{\Mean}[1]{\left< #1 \right>}
\newcommand{\half}{\frac{1}{2}}

\begin{document}
\renewcommand{\baselinestretch}{1.3}
\small\normalsize
\renewcommand{\theequation}{\arabic{section}.\arabic{equation}}
\renewcommand{\thesection}{\Roman{section}.}
\renewcommand{\thefootnote}{\fnsymbol{footnote}}
\language0

\title{Simulated-tempering approach to spin-glass simulations}
\author{Werner Kerler and Peter Rehberg}
\date{\sl Fachbereich Physik, Universit\"at Marburg, D-35032 Marburg,
Germany}
\maketitle
\begin{abstract}
After developing an appropriate iteration procedure for the determination
of the parameters, the method of simulated tempering has been successfully 
applied to the 2D Ising spin glass. The reduction of the slowing down is
comparable to that of the multicanonical algorithm. Simulated tempering
has, however, the advantages to allow full vectorization of the programs
and to provide the canonical ensemble directly.
\end{abstract}

\newpage
\renewcommand{\baselinestretch}{1.0}
\small\normalsize

\section{Introduction}
\hspace{0.35cm}
A better theoretical understanding of spin glasses is still highly desirable.
To make progress within this respect more efficient simulation algorithms are
needed. In particular at low temperatures conventional simulations suffer
{}from severe slowing down due to energy barriers. Recently Berg and Celik 
\cite{BC92} have been able to reduce the slowing down considerably by applying
the multicanonical method \cite{BN91}. Shortcomings of this method are that
the computer programs cannot be vectorized and that additional calculations
(with complicated error estimates) are needed to obtain the canonical ensemble.

An alternative is offered by the method of simulated tempering which has been
introduced by Marinari and Parisi \cite{MP92} for the random-field Ising model.
It works by associating a set of inverse temperatures to an additional 
dynamical variable of the Monte Carlo scheme. Because then the system is warmed
up from time to time one gets a chance to avoid the effective ergodicity 
breaking \cite{FH91} observed in conventional approaches. By this method at 
each of the temperatures one gets the canonical ensemble directly and there 
is nothing that prevents vectorization of the programs.

In these simulations involving the joint probability distribution of the
spins and of the additional variable one has the freedom to choose the set of
temperatures and the probability distribution of the additional variable. To 
fix these quantities in an optimal way is crucial for the efficiency of the 
method. In the first application to the random-field Ising model on 
small lattices this appeared to be not very demanding \cite{MP92}. However,
for spin glasses, in particular for larger lattices and low temperatures,
we find that the appropriate determination of these parameters is far from 
straightforward.

In the present paper we discuss the related issues in detail and develop a
systematic procedure which allows to fix the respective parameters in an 
appropriate way. We apply our method to the 2D Ising spin glass and show that 
an efficient algorithm is obtained.

We use the Edwards-Anderson Hamiltonian with two replicas,
\be
H(s, t) = - \sum_{\langle i, j \rangle} J_{ij} (s_i s_j + t_i t_j) \quad ,
\ee
where the sum goes over nearest neighbors, and $s_i=\pm 1$, $t_i=\pm 1$. 
Thus we are able to measure the order parameter \cite{FH91,BY86,MPV87}
\be
q = \sum_i s_i t_i     \quad   .
\label{ord}
\ee
We investigate samples for which $J_{ij}$ is put equal to $+1$ and $-1$ 
with equal probability. 

To each value of the additional dynamical variable $n$ of simulated tempering,
which takes the values $1,2,\ldots,N$, an inverse temperature $\beta_n$ 
with $\beta_1 < \beta_2 < \ldots < \beta_N$ is associated. The partition 
function of the joint system then becomes
\be
Z = \sum_{n, s, t} \exp \left( - \beta_n H(s, t) + g_n \right) \quad .
\ee
The parameters $g_n$ reflect the freedom in the choice of the probability 
distribution $p(n)$ of the variable $n$. 

Two steps are necessary in the calculations. In the first step the parameters 
$g_n$ and $\beta_n$ are determined. Then in the second step, using these 
parameters, the simulations are performed and the physical observables are 
measured. The dynamical variables $s_i$, $t_i$, $n$ are all updated by standard
Metropolis~\cite{Mal53} techniques. Slowing down is measured in terms of the 
ergodicity time $\tau_E$, which is defined as the mean time the system needs to
travel from $n=1$ to $N$ and back.

In Sec.~II the iteration procedures for the determination of the parameters
is described. Sec.~III contains a discussion of the  properties of this 
procedure. In Sec.~IV some numerical results are presented.

\section{Iteration procedure}
\setcounter{equation}{0}
\hspace{0.35cm}
To choose $p(n)$ constant as proposed in Ref.~\cite{MP92}, i.e.~to visit
the states $n$ with equal probability, appears reasonable. In a similar
approach \cite{KW93}, in which the additional dynamical variables is the 
number of components in the Potts model, it has been tested that such a
choice is optimal. Constant $p(n)$ in terms of the $g_n$ means that
\be
g_n = - \ln \tilde Z (\beta_n)  \label{GFree}
\ee
with the spin-glass partition function 
\be
 \tilde Z (\beta) = \sum_{s,t} \exp \left( -\beta H(s, t) \right) \quad .
\ee
It should be noted that $g_n$ depends on $\beta_n$ only.

Our choice here is to come sufficiently close to (\ref{GFree}). Because of 
the exponential dependence of $p(n)$ on $g_n$ the deviations from
(\ref{GFree}) must be small in order that all states $n$ are appropriately 
reached. We use two methods to calculate the $g_n$. The first one is to replace
$g_n$ by $g_n - \ln (Np(n))$ in subsequent iterations such that one ultimately
arrives at constant $p(n)$. The second method is to apply reweighting 
\cite{FS88} of the obtained distribution to get the ratio of the $\tilde{Z}$ 
at neighboring $n$ and thus by (\ref{GFree}) the differences of the respective 
new $g_n$. The new $g_n$ then are calculated from these differences.

To get hints on a reasonable choice of the $\beta_n$ one observes 
that the mean value of the logarithm of the acceptance probability
$\exp(-\Delta \beta H + \Delta g)$ to order $(\Delta \beta)^3$ is
\be
\left( \Mean{H^2} - \Mean{H}^2 \right) \left(\Delta \beta\right)^2 \quad .
\label{U}
\ee
On this basis one may require \cite{MP92} to choose $\beta_n$ in such a way that
(\ref{U}) gets as constant as possible. Then, due to $\Mean{H^2} - \Mean{H}^2 
\sim L^2$ at fixed $\beta$, it follows that the number $N$ of temperature 
values should scale like $N \sim L$ on the $L\times L$ lattice. Further, this 
indicates that at larger
$\beta$, where $\Mean{H^2} - \Mean{H}^2$ is small, the distances between the
$\beta$--values should increase. Unfortunately at low temperatures (large 
$\beta$) the requirement to make (\ref{U}) constant cannot be used to determine
$\Delta \beta$ because  the measurements are likely to arrive at 
$\Mean{H^2} - \Mean{H}^2 = 0$, which would result in a breakdown of an 
iterative determination of the $\beta_n$. 
 
We propose to calculate the $\beta_n$ by using a map which has an attractive 
fixed point. The map which we have constructed and applied has the property
that if the fixed point is reached in a sequence of mappings the effective stay
time $\tau_n^{eff}$ at a state $n$ gets constant. This quantity is 
defined by
\be
\tau_n^{eff} = \left\{
     \begin{array}{lll}
     \half \tau_n & \quad ,\quad & n = 1, N \\
     \tau_n       & \quad ,\quad & 2 \le n \le N-1
     \end{array} \right. \label{TauEff} 
\ee
where the stay time $\tau_n$ is the mean time which elapses between entering 
state $n$ and leaving it. To perform the map we first compute 
the auxiliary quantities
      \ba
      a_n &=& (\beta_{n+1}-\beta_n) / (\tau_{n+1}^{eff} + \tau_{n}^{eff}),
              \quad n = 1, \ldots, N-1 \quad ,   \\
      c   &=& \sum_{n=1}^{N-1} a_n \qquad . \label{CSum}
      \ea
Then we get new values $\beta_n'$ by
      \ba
      \beta_1' &=& \beta_1 \quad ,\\
      \beta_n' &=& \beta_{n-1}' + a_{n-1} \frac{\beta_n-\beta_{n-1}}{c},
                   \qquad n = 2, \ldots, N \quad . \label{BNeu}
      \ea
Since $\beta_1$ and $\beta_N$ are mapped to themselves the region of 
$\beta$ covered does not change. The fixed point gets attractive 
because a large stay time implies a smaller $\beta$--difference and thus
a smaller stay time in the next iteration step. The numerical results
indicate that by the sequence of mappings also (\ref{U}) gets very
close to a constant value.

Our procedure, which determines the parameters in an iterative way, is started 
with equidistant $\beta_n$ and with $g_n$ estimated by the relation \cite{WS88} 
$g_n=-L^2\left( a\beta_n + b\exp(-k\beta)\right) +c$. Each iteration involves 
four steps: 1) An appropriate number (typically some multiple of the expected 
ergodicity time) of Monte Carlo sweeps are performed to obtain the necessary 
data, as e.g. $p(n)$ and $\tau_n$. 2) New $g_n$ are computed by one of the two
methodes indicated above. 3) The map described above is used 
to obtain new $\beta_n$. 4) New $g_n$ which are related to these $\beta_n$ are 
determined by spline interpolation \cite{NuRe} (this step is dictated by the 
fact that $g_n$ is a function of $\beta_n$). Criteria for the termination of
the iteration procedure are discussed in Sec.~III.

\section{Properties of the procedure}
\setcounter{equation}{0}
\hspace{0.35cm}
To study the impact of the number $N$ of $\beta$--values, we have measured the 
ergodicity time as a function of $N$. Fig.~1 shows the respective results for 
lattices
$12\times 12$ and $24\times 24$. These results suggest to avoid the steep 
increase at small $N$ by choosing $N$ slightly above the minimum. Thus we put 
$N = 1.25 L$ in our simulations. This is in accordance with the remark in 
Sec.~II that one should have $N \sim L$ and it turns out to be appropriate 
on all lattices considered.

On smaller lattices we have used the first method to get the new $\beta_n$ 
described in Sec.~II. On large lattices the second one involving reweighting, 
which performs better for larger $N$, has been applied.

For lattices with $L \ge 24$ we find effects caused by the occurrence of
of me\-ta\-sta\-ble spin-glass states. Since $g_n=\beta_n f_n$, where 
$f_n$ is the free energy, one should have
\be
\frac{\Delta g}{\Delta\beta} = H_0 \quad \mbox{ for } \quad \beta \gg 1 \quad ,
\ee
where $H_0$ is the ground state energy. In our calculations 
$(g_N-g_{N-1})/(\beta_N-\beta_{N-1})$ turns out to be, in fact, with good 
accuracy an integer number. However, one actually relies on the lowest 
energy which has been reached in the Monte Carlo sweeps, which might be larger 
than $H_0$ and thus lead to a wrong determination of the parameters. 

To see which serious trouble this can cause suppose that 
\be
\frac{g_N - g_{N-1}}{\beta_N - \beta_{N-1}} =  H_0 + \Delta H 
\label{Del}
\ee
with $\Delta H > 0$ (where $\Delta H$ can take the values $4,8,\ldots$).
The probability to get from state $N$ (related to the largest $\beta$) to state
$N-1$ is min$(p_A,1)$ with $p_A=\exp\left(-(\beta_{N-1}- \beta_N)H+
(g_{N-1}-g_N)\right)$ which after inserting (\ref{Del}) becomes
\be
p_A=\exp \left( (\beta_N - \beta_{N-1})( H -H_0-\Delta H) \right) \quad .
\label{D}
\ee
{}From (\ref{D}) it is seen that if in the course of the simulations a true 
ground state with $H=H_0$ is reached, the 
probability for the transition from $N$ to $N-1$ becomes extremely small, 
i.e.~the system gets trapped at low temperature.

Therefore special care is needed to avoid a wrong determination of type 
(\ref{Del}) of the parameters. For this reason we have done a large number of
iterations in each of which we have determined the set of parameters 
and an estimate of the ergodicity time $\tau_E$. Out of these sets 
we have selected the set of parameters associated to the smallest 
$\tau_E$ which in addition has satisfied the requirement 
\be
\left| \frac{g_{N}-g_{N-1}}{\beta_N-\beta_{N-1}} - H_0 \right| < \epsilon
\label{ECond} \quad .
\ee
In (\ref{ECond}) for $H_0$ the lowest energy reached in all previous iterations
has been inserted, which has turned out to be a workable concept. Because, 
as already pointed out, for $\Delta g / \Delta \beta$ we find integers 
with good accuracy, $\epsilon = 0.1$ is a reasonable bound. The iterations
have been terminated if after an appropriate time no set of parameters with
still lower $\tau_E$ and respecting (\ref{Del}) has occurred.

On the largest lattices which we have considered to reach sufficient accuracy 
of the parameters in the iterations has turned out to be cumbersome.  If the
parameters used in a simulation are not accurate enough, the sweeps happen to 
get restricted to a subinterval of the $n$-interval which spoils the 
calculation. 

\section{Numerical results}
\setcounter{equation}{0}
\hspace{0.35cm}
Our simulations have been performed on 2D lattices with periodic boundary
conditions of sizes $L=4$, $12$, $24$,
$32$, $48$ for ten samples of $J_{ij}$-configurations in each case. This
has been done for $0.3 \le \beta \le 3.5$ putting $N = 1.25 L$. The 
ground-state energy density
\be
E_0 = \frac{1}{L^2} \min_s \left( - \sum_{\langle ij \rangle} J_{ij} s_i s_j
                           \right) \quad ,
\ee
the distribution $P(q)$ of the order parameter (\ref{ord}), and the ergodicity 
time $\tau_E$ have been determined. From the obtained data the moments 
$\Mean{q^2}$ and $\Mean{q^4}$ of $P(q)$ and the Binder parameter
\be
B_q = \half \left(3 - \frac{\Mean{q^4}}{\Mean{q^2}^2} \right) \quad .
\ee
have been calculated. The results of this for $\beta = 3.5$ are presented in
Tables 1 -- 5. For the samples the errors of the listed quantities are 
statistical ones including the effect of autocorelations (the $E_0$ are
exact). The errors of the mean values are the ones derived from the 
fluctuations of the sample values.

Our results for the mean value of the ground-state energy within errors agree
with the numbers of other authors \cite{BC92,WS88,CM83}. The results for 
$\Mean{q^2}$ and $B_q$ are compatible with those given by Berg and Celik
\cite{BC92}, though they are generally slightly larger. A possible reason for 
this is that we have measured at a lower temperature than they did. From our
tables it is seen that there are strong dependences on the particular sample
as is expected because of the lack of self-averaging~\cite{FH91,BY86,MPV87}.

Figs.~2 and 3 show typical results for the distribution $P(q)$. The invariance 
of the Hamiltonian under the transformation $s_i \to s_i$, $t_i \to -t_i$ 
requires $P(q)=P(-q)$ as is observed. The scaling law \cite{BY88} 
$P(q) = L^{0.1} \bar P (qL^{0.1})$
with an universal function $\bar P$ was verified (due to the low number of 
samples with rather large errors).

The dependence of the ergodicity time on the lattice size $L$ is depicted in 
Fig.~4 and compared to the multicanonical data given by Berg and Celik 
\cite{BC92}. We get the dependence
\be
\tau_E \sim L^{4.27(8)} \quad .
\label{tE}
\ee
A fit of the form $\tau_E \sim \exp(kL)$ is not possible. Our ergodicity
times are comparable with those of Ref.~\cite{BC92} (where the dynamical
criticial exponent 4.4(3)) is obtained). However, because our computer 
programs can be fully vectorized in terms of CPU times we gain
a large factor.

\section*{Acknowledgements}
\hspace{0.35cm}
This work has been supported in part by the Deutsche Forschungsgemeinschaft 
through grants Ke 250/7-1 and Ke 250/7-2. The computations have been done on 
the SNI 400/40 of the Universities of Hessen at Darmstadt and on the Convex 
C230 of Marburg University.

\newpage
\renewcommand{\baselinestretch}{1.2}

\newpage
\renewcommand{\baselinestretch}{1.1}
\small\normalsize

\begin{center}

{\bf Table I}

Numerical results for $4 \times 4$ lattice.

\begin{tabular}{|c||c|c|c|c|c|}
\hline
 \# & $\tau_E$   & $E_0$  &$\langle q^2\rangle$&$\langle q^4\rangle$& $B_q$ \\
\hline \hline
 1 & 70.53(8) & -1.125 & 0.2419(6) & 0.1247(5) & 0.435(9) \\
 2 & 92.00(11) & -1.500 & 0.88256(29) & 0.7927(5) & 0.9912(7) \\
 3 & 99.84(13) & -1.375 & 0.7986(4) & 0.6642(7) & 0.9793(10) \\
 4 & 78.39(9) & -1.250 & 0.4477(6) & 0.2767(6) & 0.8099(35) \\
 5 & 64.05(7) & -1.000 & 0.1746(4) & 0.07206(30) & 0.318(11) \\
 6 & 86.01(10) & -1.750 & 1.00000(1) & 1.00000(1) & 1.0000(1) \\
 7 & 85.88(10) & -1.375 & 0.6620(6) & 0.4923(7) & 0.9383(18) \\
 8 & 212.2(4) & -1.250 & 0.99983(4) & 0.99979(5) & 1.0000(1) \\
 9 & 89.02(11) & -1.375 & 0.6956(4) & 0.5174(6) & 0.9653(13) \\
 10 & 100.28(13) & -1.250 & 0.5237(10) & 0.4372(10) & 0.703(5) \\
\hline \hline
 Mean & 97(13) & -1.33(7) & 0.64(9) & 0.54(10)  & 0.81(8) \\ 
\hline
\end{tabular}

\vspace{2.5cm}

{\bf Table II}

Numerical results for $12 \times 12$ lattice.

\begin{tabular}{|c||c|c|c|c|c|}
\hline
 \# & $\tau_E$  & $E_0$   &$\langle q^2\rangle$&$\langle q^4\rangle$& $B_q$ \\
\hline \hline
 1 & 1689(14) & -1.3611 & 0.351(4) & 0.1660(28) & 0.826(15) \\
 2 & 4576(61) & -1.3611 & 0.458(5) & 0.248(5) & 0.911(25) \\
 3 & 2818(30) & -1.4444 & 0.778(2) & 0.610(2) & 0.996(4) \\
 4 & 3509(42) & -1.3611 & 0.481(7) & 0.311(7) & 0.83(4) \\
 5 & 8421(153) & -1.3333 & 0.592(4) & 0.360(4) & 0.986(12) \\
 6 & 1219(8) & -1.3750 & 0.2250(27) & 0.0872(17) & 0.64(4) \\
 7 & 3944(49) & -1.3472 & 0.5097(32) & 0.2793(33) & 0.962(13) \\
 8 & 1271(9) & -1.3472 & 0.2732(29) & 0.1210(19) & 0.690(30) \\
 9 & 2722(29) & -1.4861 & 0.7842(31) & 0.640(4) & 0.980(8) \\
 10 & 1360(10) & -1.4444 & 0.302(4) & 0.1656(31) & 0.59(4) \\
\hline \hline
 Mean & 3153(695) & -1.386(17) & 0.48(6) & 0.30(6) & 0.84(5) \\
\hline
\end{tabular}

\newpage

{\bf Table III}

Numerical results for $24 \times 24$ lattice.

\begin{tabular}{|c||c|c|c|c|c|}
\hline
 \# & $\tau_E$ & $E_0$ &$\langle q^2\rangle$ & $\langle q^4\rangle$ & $B_q$ \\
\hline \hline
 1 & 31963(516) & -1.3958 & 0.472(5) & 0.255(5) & 0.929(24) \\
 2 & 55208(1175) & -1.4271 & 0.430(9) & 0.225(7) & 0.89(5) \\
 3 & 184218(7176) & -1.3993 & 0.532(38) & 0.314(33) & 0.94(14) \\
 4 & 76188(1905) & -1.4028 & 0.576(3) & 0.342(4) & 0.985(11) \\
 5 & 108775(3289) & -1.4167 & 0.619(2) & 0.393(2) & 0.986(7) \\
 6 & 43150(826) & -1.3958 & 0.329(5) & 0.130(3) & 0.898(30) \\
 7 & 97243(2810) & -1.4271 & 0.524(5) & 0.304(5) & 0.946(18) \\
 8 & 192333(7852) & -1.3958 & 0.401(31) & 0.250(21) & 0.72(17) \\
 9 & 109485(3291) & -1.4063 & 0.599(6) & 0.382(7) & 0.968(20) \\
 10 & 40648(751) & -1.3889 & 0.328(8) & 0.167(6) & 0.72(7) \\
\hline \hline
 Mean & 93921(18067) & -1.406(4) & 0.481(34) & 0.276(28) & 0.900(31) \\
\hline
\end{tabular}

\vspace{2.5cm}

{\bf Table IV}

Numerical results for $32 \times 32$ lattice.

\begin{tabular}{|c||c|c|c|c|c|}
\hline
 \# &$10^{-5}\tau_E$&$E_0$&$\langle q^2\rangle$&$\langle q^4\rangle$ & $B_q$ \\
\hline \hline
 1 & 3.73(10) & -1.4082 & 0.500(15) & 0.271(14) & 0.96(6) \\
 2 & 15.0(6) & -1.3906 & 0.43(4) & 0.208(33) & 0.93(20) \\
 3 & 11.0(5) & -1.3867 & 0.348(13) & 0.177(7) & 0.77(8) \\
 4 & 1.147(16) & -1.4277 & 0.229(5) & 0.099(3) & 0.55(7) \\
 5 & 2.73(5) & -1.4121 & 0.443(6) & 0.217(5) & 0.949(28) \\
 6 & 8.74(34) & -1.4082 & 0.447(20) & 0.250(14) & 0.88(9) \\
 7 & 5.79(17) & -1.3945 & 0.658(6) & 0.438(8) & 0.994(10) \\
 8 & 3.45(8) & -1.4063 & 0.162(14) & 0.049(7) & 0.57(29) \\
 9 & 5.83(16) & -1.3965 & 0.576(11) & 0.343(10) & 0.983(34) \\
 10 & 2.60(5) & -1.4121 & 0.490(6) & 0.268(5) & 0.941(23) \\
\hline \hline
 Mean & 6.0(14) & -1.404(4) & 0.43(5) & 0.232(35) & 0.85(5) \\
\hline
\end{tabular}

\newpage

{\bf Table V}

Numerical results for $48 \times 48$ lattice.

\begin{tabular}{|c||c|c|c|c|c|}
\hline
 \# &$10^{-6}\tau_E$&$E_0$&$\langle q^2\rangle$&$\langle q^4\rangle$ & $B_q$ \\
\hline \hline
 1 & 4.83(27) & -1.3967 & 0.304(8) & 0.117(5) & 0.87(6) \\
 2 & 12.0(10) & -1.3967 & 0.054(10) & 0.004(3) & 0.74(79) \\
 3 & 5.5(3)   & -1.4115 & 0.35(1)   & 0.131(7) & 0.98(6) \\
 4 & 2.81(15) & -1.4054 & 0.143(12) & 0.021(6) & 0.99(24) \\
 5 & 2.06(11) & -1.3906 & 0.23(5) & 0.086(26) & 0.71(57) \\
 6 & 6.3(4) & -1.4063 & 0.48(22) & 0.26(19) & 0.93(92) \\
 7 & 6.02(34) & -1.4019 & 0.549(23) & 0.316(20) & 0.98(8) \\
 8 & 1.00(4) & -1.3845 & 0.370(7) & 0.146(5) & 0.97(4) \\
 9 & 9.2(7) & -1.4115 & 0.42(4) & 0.18(4) & 0.97(20) \\
 10 & 5.1(3) & -1.3976 & 0.28(4) & 0.085(27) & 0.94(35) \\
\hline \hline
 Mean & 5.5(10) & -1.4003(28) & 0.32(4) & 0.135(31) & 0.91(3) \\
\hline
\end{tabular}

\end{center}

\newpage
\renewcommand{\baselinestretch}{1.3}
\small\normalsize

\section*{Figure Captions}

\begin{tabular}{rl} 
Fig. 1. & Ergodicity time $\tau_E$ as a function of $N$ for lattices \\
        & $12 \times 12$ (dots) and $24\times 24$ (crosses).\\
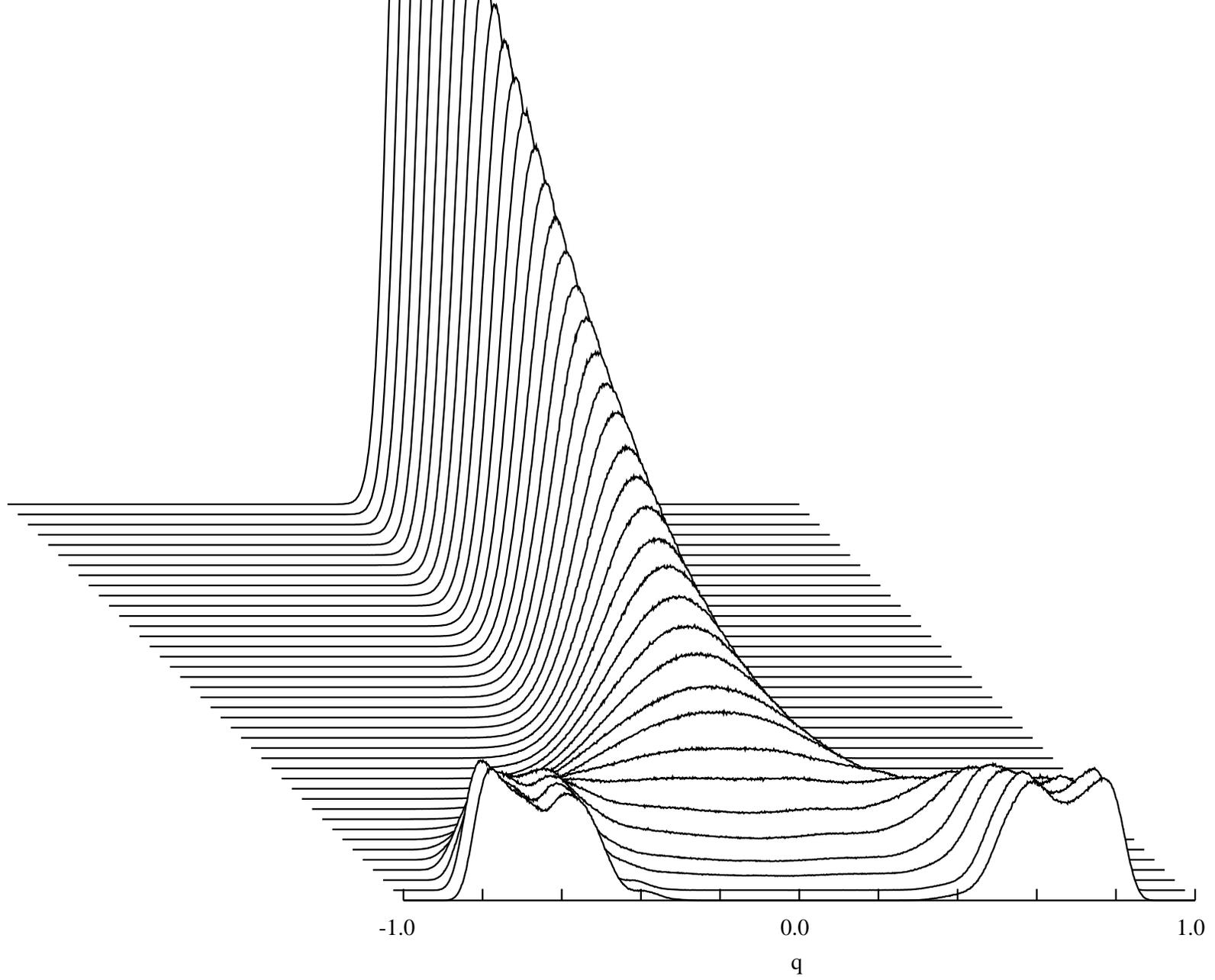
Fig. 2. & Order-parameter distribution $P(q,\beta)$ for sample 5 \\
        & on the $32 \times 32$ lattice.\\
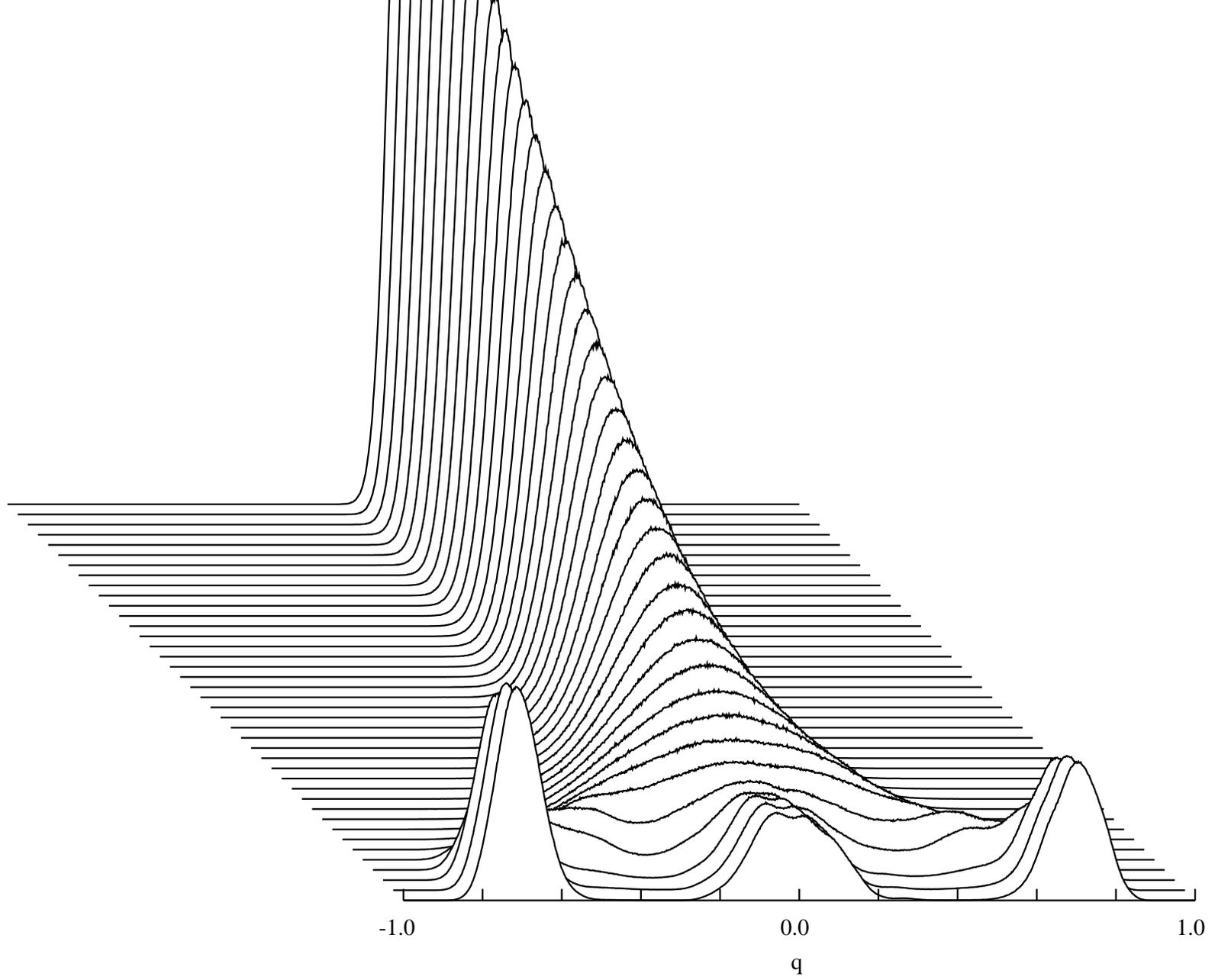
Fig. 3. & Order-parameter distribution $P(q,\beta)$ for sample 3 \\
        & on the $32 \times 32$ lattice.\\
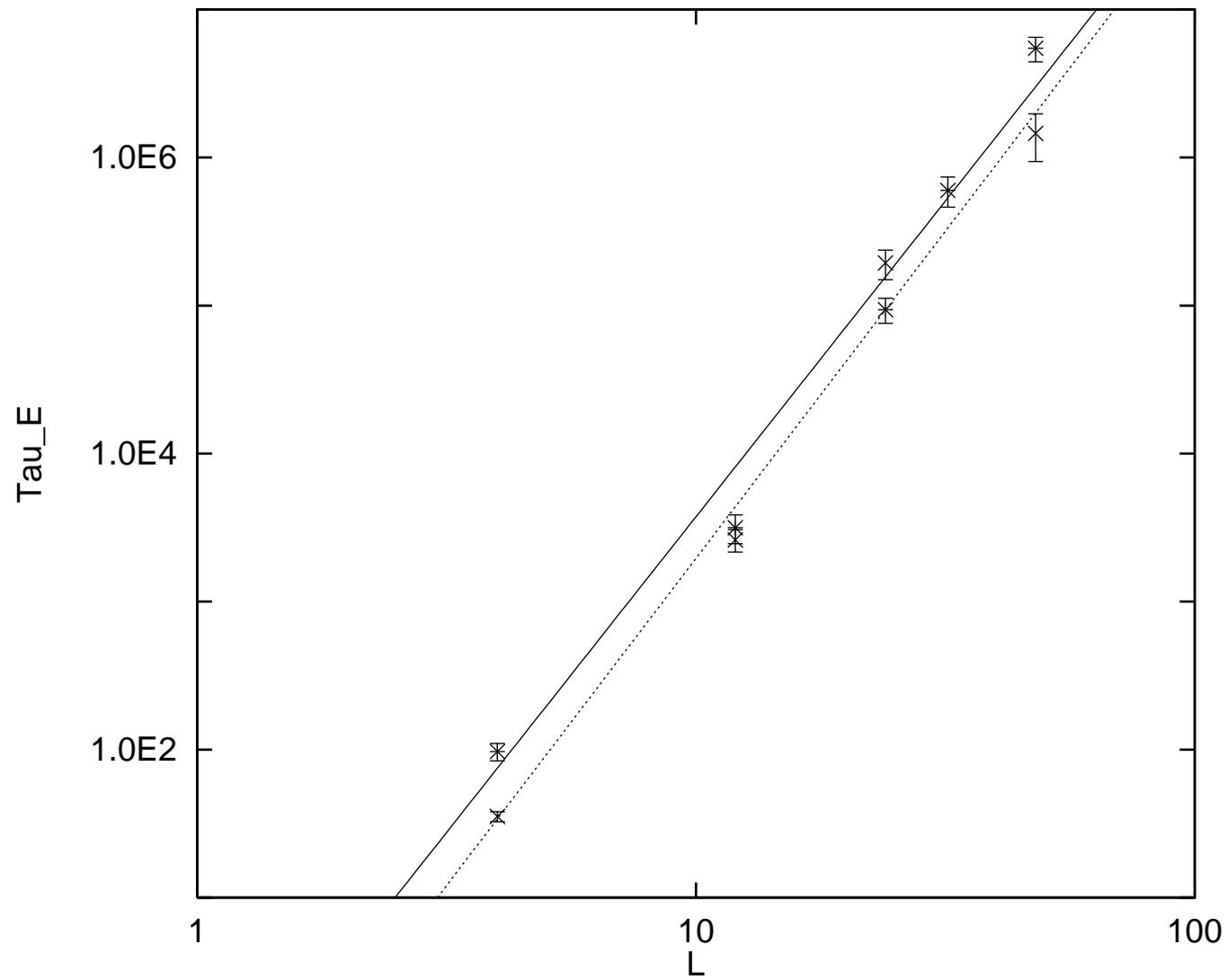
Fig. 4. & Dependence of $\tau_E$ on the lattice size $L$ for simulated\\
        & tempering (dots, fit solid line) and the multicanonical method\\
        & (crosses, fit dashed line).\\
\end{tabular}

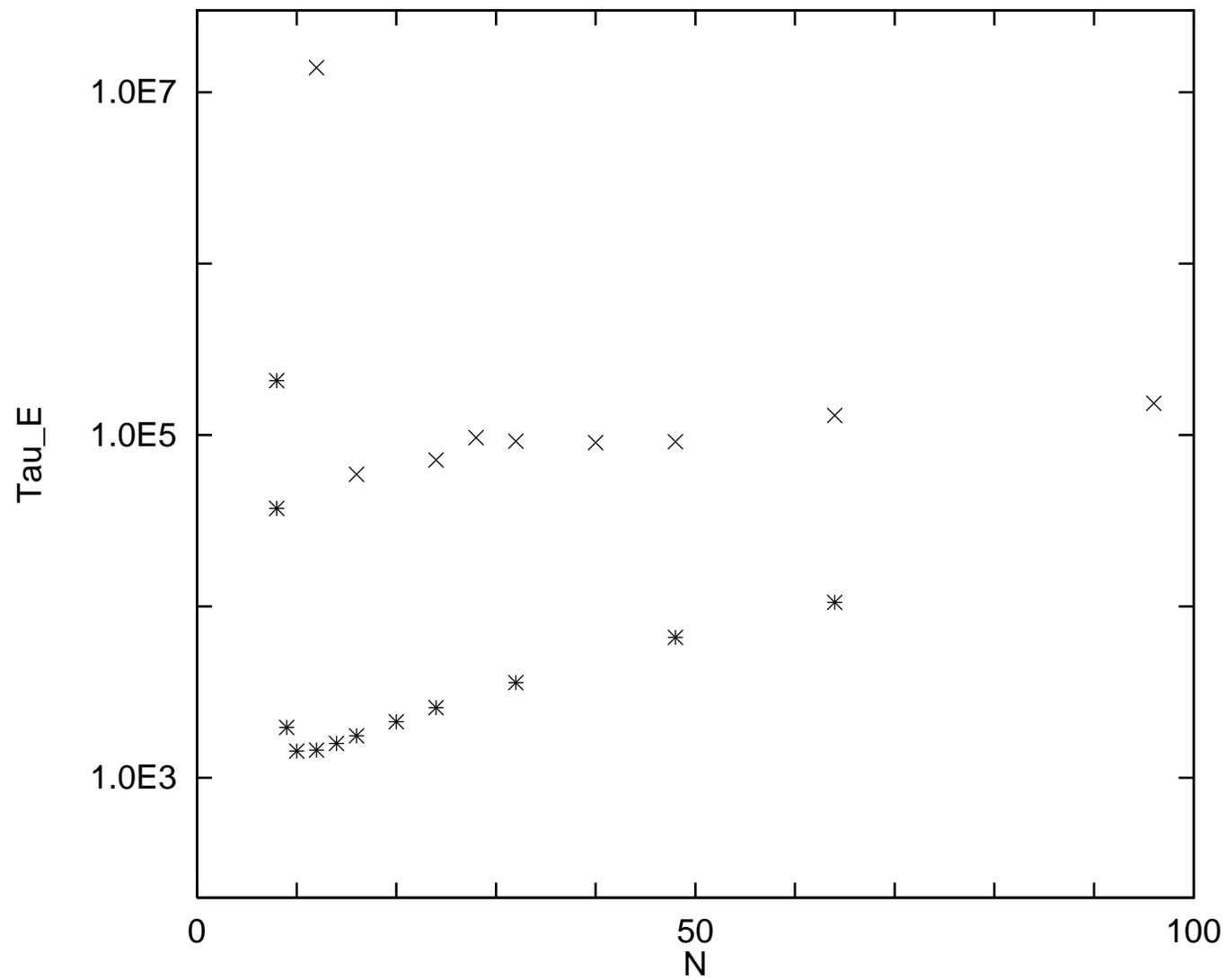
\end{document}